\documentclass[12pt]{article}
\usepackage{amssymb}
\usepackage{epsfig}

\catcode`\@=11
\def\numberbysection{\@addtoreset{equation}{section}
        \def\theequation{\thesection.\arabic{equation}}}

\def\beq{\begin{equation}}
\def\eeq{\end{equation}}
\numberbysection
\begin{document}
\begin{titlepage}
\begin{center}
\hfill  \\
\vskip 1.in {\Large \bf Hawking radiation for a scalar field conformally coupled to an $AdS$ black hole} \vskip 0.5in P. Valtancoli
\\[.2in]
{\em Dipartimento di Fisica, Polo Scientifico Universit\'a di Firenze \\
and INFN, Sezione di Firenze (Italy)\\
Via G. Sansone 1, 50019 Sesto Fiorentino, Italy}
\end{center}
\vskip .5in
\begin{abstract}
The decomposition in normal modes of a scalar field conformally coupled to an $AdS$ black hole leads to a Heun equation with simple coefficients thanks to conformal invariance. By applying the Damour-Ruffini method we can relate the critical exponent of the radial part at the horizon surface to the Hawking radiation of scalar particles.
\end{abstract}
\medskip
\end{titlepage}
\pagenumbering{arabic}
\section{Introduction}

An important application of quantum gravity has been the discover of the relation between black holes and thermodynamics. At large distance from the curvature singularity of the black hole, the gravitational effects are so weak that calculations can be based on the technique of quantum field theory on curved space.

In this way Hawking in 1974 proved that black holes can emit any kind of particle ( transforming a pure state into a mixed state ), similarly to the black-body radiation \cite{1}.  The Hawking radiation looses all the information about the black hole interior, apart from essential parameters like mass, angular momentum and charge ( the so called no-hair theorem ).

Afterwards many alternative methods have been proposed for a better understanding of this phenomenon:

i) the tunneling method; the Hawking radiation can be viewed as a tunneling process, where the barrier is created by the tunneling particle itself. To calculate this process, related to the imaginary part of the action,
two methods have been developed, one known as null-geodesic method ( Parikh and Wilczek \cite{2} ) and the other as Hamilton-Jacobi method ( Angheben and others \cite{3}).

ii) the classical Damour-Ruffini method \cite{4}, that will be reviewed in the present article. To apply this method it is necessary studying the classical solutions of a scalar field in the background of a black hole.

Our contribution is deriving the exact solution of the Klein-Gordon equation for a scalar field conformally coupled
to an $AdS$ black hole in terms of known functions. In particular the radial part is solved by a Heun function with very simple coefficients. An essential role for simplifying the solution is played by the conformally invariant  coupling with the gravitational field.

In the $AdS$ case we must avoid introducing an explicit mass term for the scalar field ( breaking conformal invariance ), because there is a spontaneous generation of a mass term from the coupling between the $AdS$ curvature and the scalar field. Thanks to the Damour-Ruffini method one can verify that the Hawking radiation arises also in this case.

In any case the exact solution of the wave function in presence of an $AdS$ black hole may be useful for a better
understanding of the gauge/gravity duality ( the $ AdS/CFT $ correspondence ) that in recent years has received
a lot of attention in literature.

\section{Properties of the $AdS$ black hole}

The $AdS$ black hole metric can be expressed in the following form:

\beq ds^2 \ = \ \frac{ Q(r) }{ r^2 } \ dt^2 \ - \ r^2 \left[ \ \frac{ dr^2 }{ Q(r) } \ + \ d \theta^2 \ + \ sin^2 \theta \ d \phi^2 \right] \label{21}\eeq

where

\beq Q(r) \ = \ \frac{ |\Lambda| }{ 3 } \ r \ \left( r^3 \ + \ \frac{ 3 }{ |\Lambda| } \ r \ - \ \frac{ 6m }{ |\Lambda| } \ \right) \label{22}\eeq

This metric satisfies the Einstein equations with negative cosmological constant outside the singularity $r=0$ and in
particular

\beq R \ = \ - \ 4 \  |\Lambda| \label{23}\eeq

The non trivial roots of $Q(r)$ , that are contained in the third degree polynomial equation, can be computed with
the Cardano formula:

\begin{eqnarray}
r_1 & = & \frac{ 1 }{ \sqrt{ |\Lambda| }} \ ( \alpha - \alpha^{-1} ) \ = \ \frac{ 6m }{ 1 + \alpha^2 + \alpha^{-2} } \ < \ 2m \nonumber \\
r_2 & = & - \frac{1}{2} \ r_1 \ + \ i \sqrt{ \frac{3}{|\Lambda|} } \left( \frac{ \alpha + \alpha^{-1} }{ 2 } \right)
\nonumber \\
r_3 & = & - \frac{1}{2} \ r_1 \ - \ i \sqrt{ \frac{3}{|\Lambda|} } \left( \frac{ \alpha + \alpha^{-1} }{ 2 } \right)
\label{24}\end{eqnarray}

where

\beq \alpha \ = \ {( \ 3 m \sqrt{|\Lambda|} \ + \ \sqrt{ 1 \ + \ 9 m^2 |\Lambda| } \ )}^\frac{1}{3} \ > \ 1
\label{25}\eeq

In the limit $\Lambda \rightarrow 0$ the complex roots $r_2$ and $r_3$ go to infinity and they decouple from the solution, while $r_1$, the only positive real root, defines the event horizon.

The gravitational acceleration on the black hole horizon surface $r_1$ and the Hawking temperature are given by:

\begin{eqnarray}
k_1 & = & \frac{ Q'(r_1) }{ 2 r^2_1 } \ = \ \frac{ |\Lambda| }{ 6 } \ \frac{(r_1-r_2)(r_1-r_3)}{ r_1 } \ = \
\frac{ 1 \ + \ |\Lambda| r^2_1 }{ 2 r_1 } =
\nonumber \\
 & = &  \frac{1 \ + \ \alpha^4 \ + \ \alpha^{-4} }{12 \ m} \ > \ \frac{ 1 }{ 4 \ m  } \nonumber \\
 T_1  & = & \frac{ k_1 }{ 2 \pi } \label{26}
\end{eqnarray}

\section{Conformal invariance}

There are many ways to couple a scalar field to gravitation; they can be parameterized by $\epsilon$ in the following action:

\beq S \ = \ \int \ d^D x \ \sqrt{|g|} \ \frac{ 1 }{ 2 } \ \left( \ g^{\mu\nu} \ \partial_{\mu} \phi \
\partial_{\nu} \phi \ - \ \epsilon \ R \ \phi^2 \ \right) \label{31}\eeq

The corresponding equation of motion is

\beq ( \ \Box \ + \ \epsilon \ R \ ) \ \phi \ = \ 0 \ \ \ \ \ \ \ \ \ \Box \ = \ |g|^{ - \frac{ 1 }{ 2 } } \
\partial_\mu \ ( \ |g|^{ \frac{ 1 }{ 2 } } \ g^{ \mu \nu } \ \partial_\nu \ ) \label{32}\eeq

Two cases are particularly relevant:

i) $\epsilon \ = \ 0$  minimal coupling;

ii) $\epsilon \ = \ \frac{ D-2 }{ 4 (D-1) }$  conformal invariant coupling. In particular
we will discuss in the following the case $\epsilon =  \frac{ 1 }{ 6 } $ \ $( D = 4)$, where the action (\ref{31}) is invariant under the following conformal transformations

\begin{eqnarray}
\tilde{g}_{\mu\nu} & = & \Omega^2 ( x ) \ g_{\mu\nu} \nonumber \\
\tilde{R} & = & \Omega^{-2} ( x ) \ [ \ R \ + \ 2 ( D-1) \ \Box \ ln \Omega \ + \ (D-1)(D-2) \ g^{\alpha\beta} \
\partial_{\alpha} ( ln \Omega ) \ \partial_{\beta} ( ln \Omega ) \ ] \nonumber \\
\tilde{\phi} & = & \Omega^\frac{2-D}{ 2 } \ \phi \label{33}
\end{eqnarray}

In (\ref{31}) we have avoided including an explicit mass term because this breaks conformal invariance, which is important to simplify the solution.

\section{Equations of motion for the scalar field}

The equation of motion for a scalar field in the background of an $AdS$ black hole is given by ( for the Kerr-Newmann case see \cite{5} )

\begin{eqnarray}
 & \ & \frac{1}{ \sqrt{|g|} } \ \partial_\mu \ ( \ g^{\mu\nu} \sqrt{|g|} \ \partial_\nu \ ) \ \psi \ + \  \frac{1}{ 6 }
 \ R \ \psi \ = \ 0 \nonumber \\
 & \ & R \ = \ - \  4 |\Lambda| \label{41}
\end{eqnarray}

By substituting the various components of the gravitational field ( eq. (\ref{21}) )  we obtain

\begin{eqnarray}
 & \ & \partial_r^2 \psi \ + \ ( \ \partial_r ln Q(r) ) \ \partial_r \psi \ - \ \frac{ r^4 }{ Q^2 (r) } \ \partial_t^2 \psi \ - \ \nonumber \\
 & \ & \ + \ \frac{ 1 }{ Q (r) } \ \left[ \ \frac{ 1 }{ \sin \theta } \ \partial_\theta ( sin \theta  \ \partial_\theta ) \psi \ + \  \frac{ 1 }{ \sin^2 \theta } \ \frac{ \partial^2 }{ \partial \phi^2 } \psi \ + \ \frac{ 2 }{ 3 }
 \ | \Lambda |  r^2  \psi \right] \ = \ 0 \label{42}
\end{eqnarray}

This equation can be resolved by separating the variables:

\begin{eqnarray}
 & \ & \psi \ = \ R(r) \ S(\theta) \ e^{ i m \phi } \ e^{ - i \omega t} \nonumber \\
 & \ & \partial_r^2 R(r) \ + \ ( \partial_r ln Q(r) ) \ \partial_r R(r) \ + \ \frac{ \omega^2 r^4 }{ Q^2 (r) } \ R(r) \ + \ \frac{ 1 }{ Q(r) } \ [ \ \lambda \ + \ \frac{ 2 }{ 3 } \ |\Lambda| \ r^2 \ ] \ R(r) \ = \ 0 \nonumber \\
 & \ & \frac{ 1 }{ sin \theta } \ \partial_\theta (  sin \theta  \ \partial_\theta ) S \ - \ \frac{ m^2 }{ sin^2 \theta }
 \ S \ = \ \lambda \ S \label{43}
\end{eqnarray}

The angular equation is equivalent to the angular momentum equation of quantum mechanics.

In this article we will discuss in detail the solution of the radial equation. We can map the two real solutions of the equation $ Q(r) = 0 $ in the points $ 0, 1$ through the following transformation

\beq z \ = \ \frac{ r ( r_1 - r_3 ) }{ ( r - r_3 ) r_1 } \label{44} \eeq

The other two finite singularities of $ Q(r) $ are mapped by the transformation (\ref{44}) into the points

\beq r_2 \rightarrow z = \xi = \frac{ r_2 ( r_1 - r_3 ) }{ ( r_2 - r_3 ) r_1 } \ \ \ \ \ r_3 \rightarrow z = \infty \label{45} \eeq

while the singularity at the infinity $ r = \infty $ is mapped into the finite point

\beq r \ = \ \infty \ \rightarrow \ z = \eta = \frac{ r_1 - r_3 }{ r_1 } \label{46} \eeq

Thanks to the identity

\beq \left( \ \frac{ 1 }{ r }, \ \frac{ 1 }{ r - r_1 }, \ \frac{ 1 }{ r - r_2 }, \ \frac{ 1 }{ r - r_3 } \
\right) \ = \ \frac{ z - \eta }{ \eta \ r_3 } \ \left( \ \frac{ \eta }{ z }, \ \frac{ \eta - 1 }{ z - 1 },
\ \frac{ \eta - \xi }{ z - \xi }, \ 1 \right) \label{47} \eeq

we can rewrite the radial equation in the $z$ variable:

\begin{eqnarray}
  \frac{ d^2 }{ d z^2 } \ R(z) & + & \ P(z) \ \frac{ d }{ d z } R(z) \ + \ Q(z) R(z) \ = \ 0 \nonumber \\
  P(z) & = & \frac{ 1 }{ z } \ + \ \frac{ 1 }{ z - 1 } \ + \ \frac{ 1 }{ z - \xi } \ - \ \frac{ 2 }{ z - \eta }
 \nonumber \\
 Q(z) & = & \frac{ \omega^2_1 }{ ( z - 1 )^2 } \ + \ \frac{ \omega^2_2 }{ ( z - \xi )^2 } \ + \ \frac{ 2 \omega_1 \omega_2 }{ ( z - 1 ) \ ( z - \xi ) } \nonumber \\
 & + & \frac{ 2 }{ ( z - \eta )^2 } \ - \ \frac{ 4  }{ \eta \ ( z - \eta ) } \ - \ \frac{ 4  }{ \eta }
 \frac{ 1 + \xi - z }{ ( z -1 ) ( z - \xi ) } \ + \ \frac{ 2 }{ ( z -1 ) ( z - \xi ) } \nonumber \\
 & + & \frac{ 3 \lambda }{ |\Lambda| \ r_1 ( r_2 - r_3 ) } \ \frac{ 1 }{ z ( z - 1 ) ( z - \xi ) } \label{48}
\end{eqnarray}

where

\beq  \omega_1 = \frac{ 3 \ \omega }{ |\Lambda| \ \eta ( r_1 - r_2 ) } \ \ \ \ \ \ \ \ \omega_2 = - \ \xi \ \omega_1
\label{49} \eeq

Apparently this equation depends on the singularity $ z = \eta $, but if we put this differential equation into the
normal form, the whole dependence from $ z - \eta $ is removed, thanks to the conformal invariance of the coupling
with the gravitational field ( see also  \cite{6} ):

\begin{eqnarray}
R(z) & = & e^{-\frac{ 1 }{ 2 } \int P(z) \  dz} \ Z(z) \nonumber \\
\frac{ d^2 }{ dz^2 } Z(z) & + & \tilde{ Q }(z) \ Z(z) = 0 \label{410}
\end{eqnarray}

where

\begin{eqnarray}
\tilde{ Q }(z) & = & Q(z) \ - \ \frac{ 1 }{ 2 } \ \frac{ d }{ dz } P(z) \ - \ \frac{ 1 }{ 4 } \ P^2(z) = \nonumber \\
& = &  \frac{ 1 }{ 4 \ z^2 } \ + \ \frac{ 1 + 4 \ \omega^2_1}{ 4 \ ( z-1 )^2 } \ + \
 \frac{ 1 + 4 \ \omega^2_2}{ 4 \ ( z-\xi )^2 } \ + \ \frac{ 4 \ \omega_1 \omega_2 - 1 }{
2 \ (z-1) ( z-\xi ) } \nonumber \\
& + & \left( \frac{ 1 + \xi }{ 2 } \ - \ \frac{ \xi }{ \eta } \ + \ \frac{ 3 \lambda }{  |\Lambda| r_1 ( r_2 - r_3 )  } \right) \frac{ 1 }{ z ( z-1 ) ( z-\xi) } \label{411}
\end{eqnarray}

Let us notice that this equation can be connected with the known Heun equation

\beq \frac{ d^2 }{ dz^2 } w(z) \ + \ P_H(z) \ \frac{ d }{ dz } w(z) \ + \ Q_H(z) w(z)  \ = \ 0 \label{412} \eeq

where

\begin{eqnarray}
P_H(z) & = & \frac{ \gamma }{ z } \ + \ \frac{ \delta }{ z-1 } \ + \ \frac{ \epsilon }{ z-\xi } \nonumber \\
Q_H(z) & = & \frac{ \alpha \beta z - q  }{ z ( z-1 ) ( z-\xi )} \label{413}
\end{eqnarray}

After rewriting this equation in the normal form, we must identify

\begin{eqnarray}
w(z) & = & e^{-\frac{ 1 }{ 2 } \int P_H(z) \ dz} \ Z(z) \nonumber \\
\tilde{ Q }(z) & = &  \frac{ \frac{\gamma}{2} ( 1 - \frac{\gamma}{2} ) }{ z^2 } \ + \
\frac{ \frac{\delta}{2} ( 1 - \frac{\delta}{2} )}{ ( z-1 )^2 } \ + \
\frac{ \frac{\epsilon}{2} ( 1 - \frac{\epsilon}{2} )}{ ( z-\xi )^2 } \nonumber \\
& + & \frac{ [ \alpha \beta - \frac{ 1 }{ 2 } ( \gamma \delta + \gamma \epsilon + \delta \epsilon ) ]
}{ (z-1) ( z-\xi ) } \ - \  \frac{ q - \frac{ 1 }{ 2 } ( \gamma \delta \xi + \gamma \epsilon ) }{ z ( z-1 ) ( z-\xi) } \label{414} \end{eqnarray}

By adding the Heun condition

\beq \alpha \ + \ \beta \ = \ \gamma \ + \ \delta \ + \ \epsilon \ - \ 1 \label{415} \eeq

we have the following table of identifications

\begin{eqnarray}
\alpha & = & \gamma \ = \ 1 \nonumber \\
\beta & = & 1 \ + \ 2 i \ ( \ \omega_1 \ + \  \omega_2 \ ) \nonumber \\
\delta & = & 1 \ + \ 2 i \ \omega_1 \nonumber \\
\epsilon & = &  1 \ + \ 2 i \ \omega_2 \nonumber \\
q & = & \frac{ \xi }{ \eta } \ - \ \frac{ 3 \lambda }{ |\Lambda| r_1 ( r_2 - r_3 )  }
\label{416} \end{eqnarray}

\section{Hawking temperature}

Let us recall that

\beq R(z) \ = \ e^{-\frac{ 1 }{ 2 } \int P(z) \ dz} \ Z(z) \ = \ \frac{ z - \eta }{\sqrt{ z ( z - 1 )( z - \xi ) }} \
Z(z) \label{51} \eeq

Let us note that the two solutions for $R(r)$ in our case behave as $ 1/r $ for large $r$ and are both not normalizable ( in and out states ). The AdS space is therefore transparent at infinity for a conformal coupling.

The structure of the solution changes drastically in the unphysical region ( without conformal invariance ), with only one not normalizable solution. The AdS space acts now as a reflecting barrier.  

Near the singularity $ z = 1 $ the behaviour of the wave function is ( introducing the time factor )

\beq \psi(z) \ = \ c_1 e^{ - i \omega t } ( z-1 )^{ i \omega_1 } \ + \ c_2 e^{ - i \omega t } ( z-1 )^{ - i \omega_1 } \label{52} \eeq

We can relate the critical exponent $\omega_1$ ( eq. (\ref{49}) )

\beq \omega_1 \ = \ \frac{ \omega }{ 2 k_1 } \label{53} \eeq

 to $k_1$, the gravitational acceleration at the black hole horizon surface, and therefore to the Hawking temperature. At the horizon surface the ingoing and outgoing solutions are parameterized by

\begin{eqnarray}
\psi_{in} \ = \ e^{ - i \omega t } ( r-r_1 )^{ -i \frac{\omega}{2 k_1} } \nonumber \\
\psi_{out} ( r > r_1 ) \ = \ e^{ - i \omega t } ( r-r_1 )^{ i \frac{\omega}{2 k_1} } \label{54}
\end{eqnarray}

To simplify this discussion we can introduce the tortoise coordinate defined in general as

\beq d r_* \ = \ \frac{ r^2 }{ Q(r) } \ dr \label{55} \eeq

We can integrate this equation around $r \sim r_1$

\beq r_* \ = \ \frac{ 1 }{ 2 k_1 } \ ln ( r-r_1 ) \label{56} \eeq

where the wave equation reduces to

\beq \left( - \frac{\partial^2}{\partial t^2} \ + \ \frac{\partial^2}{\partial r_*^2} \right) \ \psi ( r_*, t ) \ = \ 0 \label{57} \eeq

The ingoing and outgoing wave functions are therefore given by

\beq \psi^{in} \ = \ e^{ - i \omega ( t + r_* )} \ \ \ \ \ \psi^{out} \ = \ e^{ - i \omega ( t - r_* )} \label{58} \eeq

By changing from the problematic time coordinate to the well behaved Eddington coordinate $ \nu = t + r_* $ the solutions discussed in (\ref{54})
are re-obtained

\begin{eqnarray}
\psi^{in} & = & e^{ - i \omega \nu } \nonumber \\
\psi^{out} & = & e^{ - i \omega \nu } e^{ 2 i \omega r_* } \ = \ e^{ - i \omega \nu } {( r - r_1 )}^{ i \frac{\omega}{ k_1} } \label{59} \end{eqnarray}

While the ingoing wave function is analytic the outgoing wave function has a logarithmic singularity at the horizon. We will use these solutions to investigate the Hawking radiation for scalar particles.

\section{Damour-Ruffini method}

To understand the decay rate it is necessary to build a damped part in the outgoing wave function ( Damour-Ruffini method ). The outgoing wave function is known only for $ r > r_1 $ therefore in the region outside the event horizon. A simple analytic continuation in the internal region gives:

\begin{eqnarray}
( r - r_1 )  & \rightarrow  & e^{ - i \pi } ( r_1 - r)  \nonumber \\
\psi_{out} ( r < r_1 ) & = & e^{ - i \omega \nu } \ ( r_1 - r )^{ i \frac{\omega}{ k_1} } \ e^{ \frac{\pi}{k_1} \omega } \label{61} \end{eqnarray}

Since the wave function is not analytic, the continuation produces a damping factor and the
scattering probability of the scalar wave at the event horizon is given by

\beq \Gamma_1 \ = \ \left| \frac{\psi_{out} ( r > r_1 )}{\psi_{out} ( r < r_1 )} \right|^2 \ = \ e^{ - \frac{2\pi}{k_1} \omega } \label{62}
\eeq

By introducing a normalization constant $N_\omega$, the normalization condition of the wave function fixes the behaviour of $N_\omega$ :

\beq |N_\omega|^2 \ = \ \frac{ 1 }{ e^{\frac{ 2 \pi }{ k_1 } \omega } - 1  } \ = \
\frac{ 1 }{ e^{\frac{ \hbar \omega }{ k_B T_1 } } - 1  } \label{63} \eeq

This formula implies that the Hawking radiation spectrum for the scalar particles has a thermal character analogous to the black-body spectrum with Hawking temperature $T_1$.

\section{Conclusions}

In this paper we have presented the decomposition in normal modes of a scalar field conformally coupled to an $AdS$ black hole. In particular the radial part leads to a differential equation with four finite singularity and the singularity at infinity. This scheme resembles the isomonodromy problem \cite{6}, however there is an important difference in our case. In fact one of the four finite singularities can be removed by simply choosing a conformal invariant coupling between the scalar field and the black hole. The resulting Heun equation ( with only three finite
singularities ) has very simple coefficients and the critical exponent at the horizon singularity can be directly related to the Hawking radiation.

We expect that this study can be generalized to the case of integrable $2+1$ gravity, in which case black holes exist only for negative cosmological constant. Our study may have possible connections with the $AdS/CFT$ correspondence and may be useful to study non-perturbative effects of quantum gravity.


\begin{thebibliography}{999}
\bibitem{1} S. W. Hawking, Comm. Math. Phys. {\bf 43} (1975) 199.
\bibitem{2} M. K. Parikh and F. Wilczek, Phys. Rev. Lett. {\bf 85} (2000) 5042; arXiv: hep-th/9907001.
\bibitem{3} M. Angheben, M. Nadalini, L. Vanzo and S. Zerbini, JHEP {\bf 05} (2005) 014.
\bibitem{4} T. Damour and R. Ruffini, Phys. Rev. D {\bf 14} (1976) 332.
\bibitem{5} H. S. Vieira, V. B. Bezerra, C. R. Muniz, Ann. of Phys. {\bf 350} (2014) 14; arXiv:1401.5397v3 [gr-qc].
\bibitem{6} F. Novaes and B. Carneiro da Cunha, JHEP {\bf 07} (2014) 132; arXiv:1404.5188v1 [hep-th].

\end{thebibliography}
\end{document}